# Regulating MDA-MB-231 breast cancer cell adhesion on laser-patterned surfaces with micro- and nanotopography

M. Kanidi[a)]

National Hellenic Research Foundation, Theoretical and Physical Chemistry Institute, 48 Vasileos Constantinou Ave., Athens 11635, Greece

A. Papadimitropoulou[a)]

National Hellenic Research Foundation, Theoretical and Physical Chemistry Institute, 48 Vasileos Constantinou Ave., Athens 11635, Greece

Biomedical Research Foundation of the Academy of Athens, 4 Soranou Ephessiou St., 115 27 Athens, Greece

C. Charalampous, Z. Chakim

Biomedical Research Foundation of the Academy of Athens, 4 Soranou Ephessiou St., 115 27 Athens, Greece

G. Tsekenis

National Hellenic Research Foundation, Theoretical and Physical Chemistry Institute, 48 Vasileos Constantinou Ave., Athens 11635, Greece

Biomedical Research Foundation of the Academy of Athens, 4 Soranou Ephessiou St., 115 27 Athens, Greece

A. Sinani, C. Riziotis, M. Kandyla[b)]

National Hellenic Research Foundation, Theoretical and Physical Chemistry Institute, 48 Vasileos Constantinou Ave., Athens 11635, Greece

[a)]These authors contributed equally to this work.
[b)]Electronic email: kandyla@eie.gr

Breast cancer is the most common type of cancer observed in women. Communication with the tumor microenvironment allows invading breast cancer cells, such as triple negative breast cancer cells, to adapt to specific substrates. The substrate topography modulates the cellular behavior, among other factors. A number of different materials and micro/nanofabrication techniques have been employed to develop substrates for cell culturing. Silicon-based substrates present a lot of advantages as they are amenable to a wide range of processing techniques and they permit rigorous control over the surface structure. We investigate and compare the response of triple negative breast cancer cells (MDA-MB-231) on laser-patterned silicon substrates with two different topographical scales, *i.e.*, the micro- and the nanoscale, in the absence of any other biochemical modification. We develop silicon surfaces with distinct morphological characteristics by employing two laser systems with different pulse durations (nanosecond and femtosecond) and different processing environments (vacuum, $SF_6$ gas, and water). Our findings demonstrate that surfaces with micro-topography are repellent, while surfaces with nano-topography are attractive for MDA-MB-231 cell adherence.

## I. INTRODUCTION

Breast cancer is the most common type of cancer observed in women, being the second leading cause of death [0, 2]. Among all subtypes of breast cancer, the triple negative one (Triple Negative Breast Cancer-TNBC) shows increased heterogeneity and lacks the expression of estrogen receptor (ER), progesterone receptor (PR), and human epidermal growth factor Receptor-2 (HER2) overexpression, therefore becomes very aggressive and resistant to several anti-cancer agents [3, 4]. Proliferation and migration constitute two major hallmarks of cancer, being the main factors for cancer-related deaths worldwide. The mechanisms underlying the dissemination of cancer cells to distant sites still remain poorly understood, despite the increasing evidence that have shed light to the factors affecting this procedure. Communication with the tumor microenvironment allows TNBC invading cancer cells, such as MDA-MB-231, to adapt to specific substrates in order to grow, proliferate, and disseminate to distant sites.

To elucidate the mechanisms underlying cancer cell survival, proliferation, and metastatic potential in vitro, a number of parameters should be taken into consideration such as the surface chemistry, surface stiffness, and topographic characteristics [5-8]. The ideal substrate for tumor cell culturing should be able to combine effectively mechanical properties and macro/nano scale topography in order to achieve the optimal conditions for cell survival and, most importantly, to mimic the microenvironment cells naturally reside in. This way, it can promote interactions between cells and the external environment (such as exchange of biochemical signals, oxygen supply, *etc.*), which are equivalent to the ones cells encounter in real tissues [9-12]. The substrate topography modulates the cellular behavior, as it has been shown that topographical features of different length-scales, ranging from nano- to micrometer ones, have a strong influence on cell adhesion, morphology, alignment, and contact guidance [13]. Even though it is understood that micro- and nanoscale topographic patterns resemble the extracellular environment, the topography of which constitutes a major biophysical regulator of cell behavior, to date cell reaction to topography remains not fully understood. It is anticipated that the optimum topography varies depending on the cell type, substrate material, and structure. Especially for cancer cells, the tumor microenvironment can greatly influence cell behavior and dissemination [14].

Recently, it has been shown that micro- and nanofabrication techniques have provided invaluable tools for studying various processes, such as cell proliferation [15], differentiation [16, 17], adhesion [18-22], and migration [23-30] through the introduction of topographical features on cell-culturing platforms [31]. At the same time, a number of different materials have been employed as substrates for cell culturing [32, 33]. Amongst them, silicon-based substrates present a lot of advantages as they are amenable to a wide range of processing techniques common to silicon microfabrication, thus providing access to a big variety of tailored geometries and surface patterns, while at the same time permitting rigorous control over the structure and composition of the surface [34]. Furthermore, silicon can be surface-modified and replicated in polymeric scaffolds, which, coupled with the fact that it is not toxic to cells, makes it an ideal substrate to study several cellular properties such as adhesion, morphology, proliferation, and migration [35, 36] on patterned surfaces and devices [37].

Patterned and/or bio-modified silicon has been employed to investigate the effect of topography on fibroblasts [38], neuronal [39, 40], and epithelial cells [41, 42] as well as on various cancer cell lines [43-48], including mammary tumor cells, such as MCF-7 and MDA-MB-231. MCF-7 and MDA-MB-231 cell lines are derived from the pleural effusions of a breast cancer patient with metastasis but exhibit diverse phenotypic features. MCF7 cells are ER+,PR+,HER2-, they are estrogen-dependent and express markers for the luminal epithelial phenotype. MCF-7 breast tumor cells form multicellular nodules (foci) over a confluent monolayer in the presence of estradiol (E2), they are poorly aggressive, and exhibit low metastatic ability. On the contrary, MDA-MB-231 do not express epithelial markers, they are ER-,PR-,HER2- (triple negative) and have acquired a highly aggressive mesenchymal phenotype [49, 50]. The rate of adhesion and growth of breast cancer MCF-7 cells was studied on mesoporous silicon [44], while MDA-MB-231 cells have been co-cultured with MCF10A non-tumorigenic human breast cells on microstructured silicon substrates with arrays of etched microcavities to study their adhesion pattern [51, 52], on micromachined arrays of silicon walls to study their plasticity [53], and a few other silicon microstructures. Specifically, fluorescently tagged cells were used to study the behavior of the highly invasive human breast cancer (MDA-MB-231) in co-culture with normal epithelial human breast (MCF10A) cells on isotropically etched silicon microarrays. Micro co-cultures comprised of 2-10 cells assembled rapidly inside etched cavities of 140-µm diameter. The cells occupied 97 – 100% of the etched cavities and were enriched in MDA-MB-231 cells relatively to the initial seeding density (1:10 MDA-MB-231/MCF10A) cells. Micro co-cultures with both cell lines formed in 26% of the cavities and comprised 2-10 cells per cavity. Another study with isotropically-etched concave cavities showed a quick and efficient capture of MDA-MB-231 from single-cell suspensions and supported the co-culture with normal breast cells without any modification of the surface. These findings demonstrate the capacity of etched silicon micro-arrays to act as an efficient substrate for the propagation of human breast cell cultures [51, 52]. In addition, another study has shown that various cancer cell lines applied on 3D silicon micromachined structures (SMS), with arrays of 3-µm-thick silicon walls separated by 50-µm-deep, 5-µm-wide gaps, exhibited different behavior according to their invasiveness. The more invasive cancer types gave higher fractions of cells in the gaps, indicating a direct correlation between the aggressiveness of tumor cells and their capacity to grow into the narrow gaps of the 3D-SMS [53].

In this work, we investigate and compare the response of MDA-MB-231 cells on laser-patterned silicon substrates with two different topographical scales, *i.e.*, the micro- and the nanoscale, in the absence of any other biochemical modification. We were able to develop silicon surfaces with distinct morphological characteristics by employing two laser systems with different pulse durations (nanosecond and femtosecond) and different processing environments (vacuum, $SF_6$ gas, and water). Laser processing is a single-step, maskless, tabletop method to create uniformly micro/nanopatterned surfaces over large areas, which does not require a clean room. Laser processing of silicon has been shown to generate a variety of structures at the

micro- and nanoscale by tuning the fabrication parameters, such as wavelength, pulse duration, fluence, gas or liquid environment, and the number of incident laser pulses [54-56]. Laser-patterned silicon substrates have been used for the culture of Schwann cells [57], fibroblasts [47], and rat pheochromocytoma (PC12) cells [58], but have not been applied to TNBC (MDA-MB-231) cell culturing before. Furthermore, laser patterning of surfaces allows for localized surface modification and high spatial resolution of the formed structures, thus paving the way for the design of co-culturing substrates and devices based on topographical cues. In this way, invaluable information on the proliferative and migrating behavior of breast cancer cells will be provided, which will enable the formulation of novel treatment approaches.

## II. EXPERIMENTAL

Cell culture. Triple Negative Breast Cancer cell line (MDA-MB-231) was purchased from American Type Culture Collection (ATCC, Manassas, VA, USA). Cells were cultured in DMEM containing 4.5 g/L D-Glucose (Sigma-Aldrich, cat. D6429), supplemented with 10% heat-inactivated fetal bovine serum (Biosera, cat. FB-1001/500) and penicillin (100 units/ml)/ streptomycin (100 lg/ml; Thermo Fischer Scientific, cat. 15140122) in a humidified incubator at 37℃ with 5% $CO_2$. When cells reached 75-80% confluency, they were treated with a solution of 0.25% Trypsin-EDTA to detach the cells from the surface of the plate. Trypsin was deactivated with the addition of medium and cells were centrifuged for 5 minutes at 1500 rpm. After centrifugation, cells were counted and 5000 cells were seeded on the sterilized silicon surfaces for 72-h incubation. Glass coverslips were used as a positive control for cell seeding.

Laser-structured silicon substrates. Silicon wafers ((100), thickness 500 ± 25 μm) were cleaned in an ultrasonic bath of acetone and methanol for 15 min and dried in nitrogen ($N_2$) gas flow before laser structuring. Microstructured silicon substrates were fabricated by nanosecond laser processing in a gas environment of sulfur hexafluoride ($SF_6$) or in vacuum. A pulsed Q-switched Nd:YAG laser system was used with 532 nm wavelength, 5 ns pulse duration, and 10 Hz repetition rate. Silicon wafers were placed in a vacuum chamber, which was evacuated to $9*10^{-3}$ mbar (vacuum conditions) or filled with 0.6 bar $SF_6$. The laser beam was focused on the silicon surface through a quartz window using a lens of 200 mm focal length and the vacuum chamber was raster scanned with respect to the laser beam using a computer-controlled set of *xy* translation stages. Nanostructured silicon substrates were fabricated by femtosecond laser processing in water. An ultrashort fiber-based laser system was used with 1064 nm wavelength, 370 fs pulse duration, maximum average power 1W, maximum pulse energy 1μJ, and variable pulse repetition rate up to 1 MHz. The laser beam was focused by a 4× microscope objective lens with NA = 0.10 on the surface of the silicon wafers that were placed at the bottom of a container filled with 6 ml of distilled water, which covered the surface of the wafer. The container was mounted on a computer-controlled set of *xy* linear translation stages with resolution 1 nm and maximum velocity 20 mm/s. For silicon substrates nanostructuring, a pulse energy of 0.7 μJ and a repetition rate of 10 kHz were identified as the optimum laser irradiation conditions. The substrates were raster scanned with respect to the laser beam with a scanning speed of 190 μm/s and 9 μm spacing between successive lines, thus providing a uniform nanomorphology on the surface.

Characterization. The surface morphology of laser-structured silicon substrates was investigated with field-emission scanning electron microscopy (SEM). Cellular attachment on various substrates was studied by fluorescence microscopy. As mentioned above, 5000 cells were incubated on silicon surfaces for 72 h at 37°C with 5% $CO_2$ in a humidified incubator, and washed twice with phosphate-buffered saline (PBS) to remove non-adhering cells floating in the medium. Subsequently, cells were fixed with 4% paraformaldehyde for 20 min on ice, washed 3 times with PBS permeabilized with 0.5% Triton X-100 in PBS, incubated with 5% serum or 1% BSA in PBS+0.1% Tween 20 to block non-specific binding and washed again 3 times with PBS. Finally, cells were incubated with antifade reagent containing DAPI (Vector Laboratories, Inc., CA 94010, USA) to visualize the nuclei and sealed with nail polish to prevent drying and movement under the microscope. Cell imaging was performed using Leica DMRA2 and Hamamatsu ORCA-Flash 4.0 V2. Image processing and cell counting were performed using ImageJ. For each surface (silicon and glass coverslip), each experiment was repeated three times.

Statistical analysis. Data were analyzed with the Prism Software (GraphPad) and the tests used were the unpaired Student t test and One Way Anova. The p values with $p < 0.05$ were considered significant (*$p < 0.05$, ****$p<0.0001$) and data were expressed as mean ± SEM (standard error of the mean) from three independent experiments (see supplementary material at [URL will be inserted by AIP Publishing]).

## III. RESULTS AND DISCUSSION

### A. MDA-MB-231 cell culturing on microstructured substrates

Figure 1a shows a SEM micrograph of a microstructured silicon area, surrounded by unprocessed flat silicon. The microstructured area is shown in higher magnification in Fig. 1b, at side (45°) view. A quasi-ordered array of conical microspikes covers the structured surface. The average height of the spikes is 35 ± 4 μm, the average half-height width 11 ± 2 μm, and the average distance between neighboring spikes 16 ± 2 μm center-to-center, which results in a spike density of $8.6*10^5$ spikes/cm$^2$. This morphology was obtained when irradiating silicon in SF$_6$ ambient with a fluence of 1 J/cm$^2$ and translating the silicon surface with respect to the laser beam with a speed such that each spot on the silicon surface was irradiated by 1000 laser pulses [59, 60]. At this fluence, this is the result of melting, ablation, and interference effects that occur upon nanosecond laser irradiation of silicon in SF$_6$ environment [55, 56].

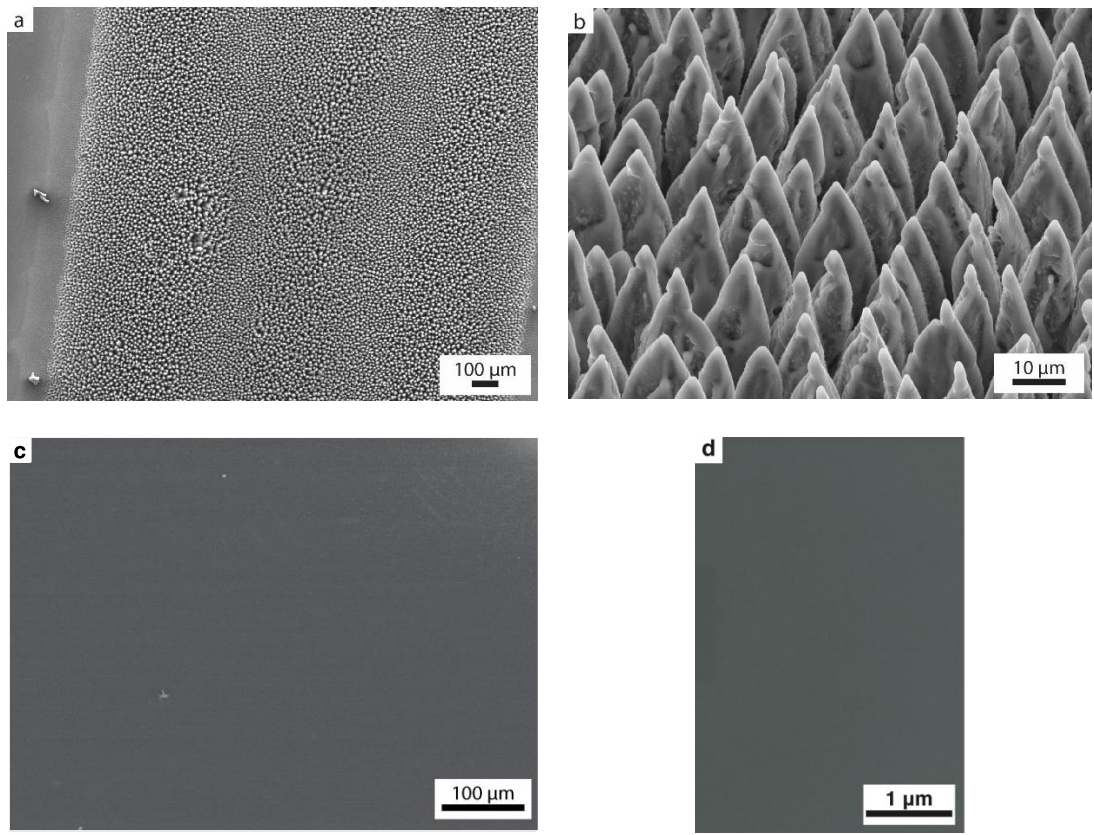

FIG. 1. Scanning electron micrographs (SEM) at (a) top view and (b) side view (45º) of laser-microstructured silicon with nanosecond laser beam in SF$_6$ (fluence 1 J/cm$^2$). (c) SEM image of flat silicon and (d) glass coverslip.

The results of MDA-MB-231 cell attachment on the microstructured silicon surface, relative to a flat silicon surface (Fig. 1c) and a control glass coverslip (Fig. 1d), are shown in Fig. 2. Figure 2a shows a fluorescent microscope image of cell attachment on the coverslip. We observe that cells were uniformly spread throughout the surface, which was identical to the behavior of the cells on the flat silicon surface (Fig. 2b). On the contrary, surface microstructuring, through the introduction of anisotropic microcones, seemed to act as a repellent for TNBC cells (Fig. 2c). For a direct comparison, Fig. 2d shows the results of cell adherence on the border between a flat (right) and a microstructured (left) silicon surface. We clearly observe the flat silicon side is cell-adhesive and the microstructured silicon side is cell-repellent. These observations were further confirmed by quantitative analysis, which showed 4028 ± 110.7 (N=3) cells on the glass coverslip (control surface), 4639 ± 196.8 (N=3) cells on flat silicon, and 1498 ± 69.04 (N=3) cells on microstructured silicon. When the cells on microstructured silicon were compared to those on flat silicon on the same sample (Fig. 2d), there was a significant difference on the adhesion of cells between them (222 ± 18.19 (N=3) to 1797 ± 99.51 (N=3), respectively).

Similar structures have been employed for the growth of neuronal cells, where they have been shown to promote and direct neuronal outgrowth [61], however, in the case of TNBC MDA-MB-231 cells, they hinder cell attachment. This is probably due to the inherent hydrophobicity of these surfaces [62], which act as a barrier for migratory cells, a finding which is further confirmed by other studies where microstructures and their effect on cell behavior were investigated [26]. Laser-structured silicon microspikes in $SF_6$ have been shown to be covered by a surface layer with a sulfur concentration of 0.5 – 0.7% [63, 64]. Additionally, because the samples used in this work were exposed to air for several days before being used as cell culture substrates, a native oxide layer was also formed on the surface, thus the MDA-MB-231 cells interact with the oxidized surface. Therefore, the presence of sulfur underneath the oxide layer is not expected to affect the results of cell behavior. Indeed, preliminary data for MDA-MB-231 cell culture on nanosecond laser-structured silicon microspikes in vacuum (not shown here) also demonstrated a similar cell-repellent behavior.

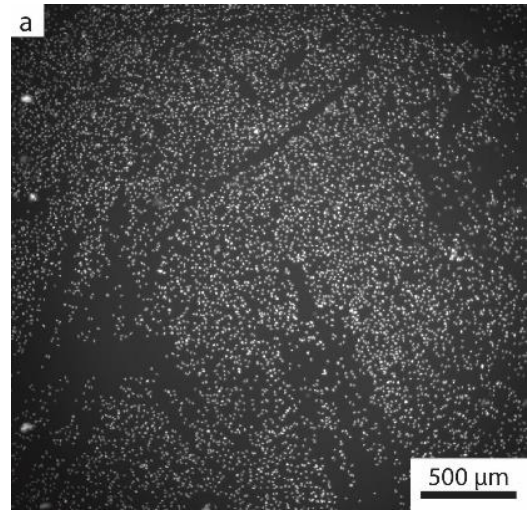

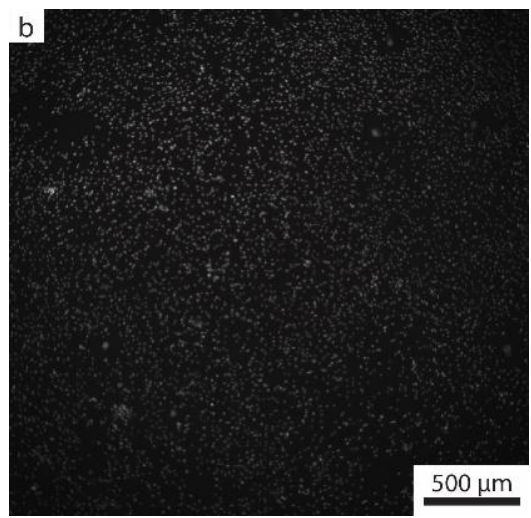

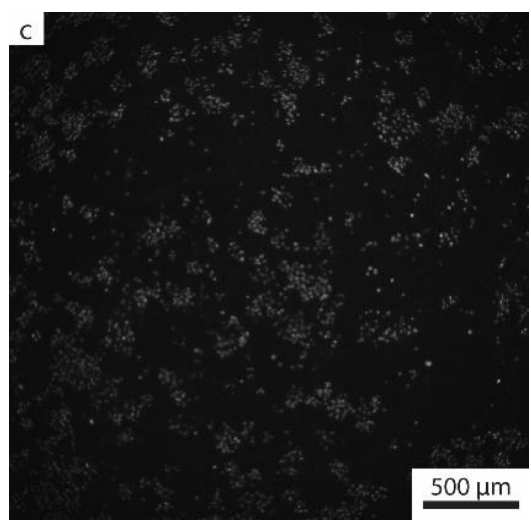

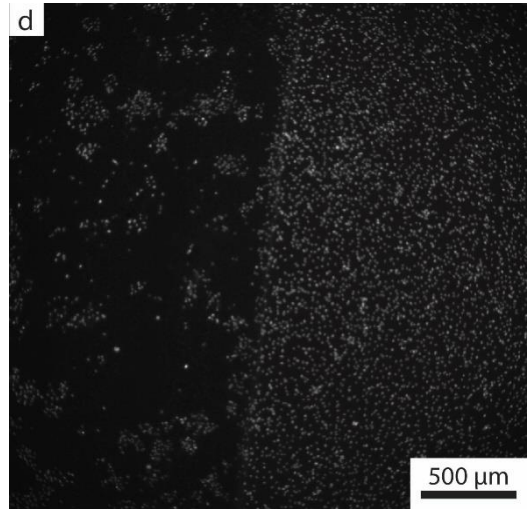

FIG. 2. Fluorescent micrographs of MDA-MB-231 cells on (a) glass coverslip, (b) flat silicon, (c) microstructured silicon as in Fig. 1, and (d) border between microstructured silicon (left side) and flat silicon (right side). Cell nuclei were stained with DAPI.

Increasing the nanosecond laser fluence, we created crater structures on the surface of silicon. Figure 3 shows an area on the silicon surface, which has been irradiated by 1000 pulses with a laser fluence of 2 J/cm$^2$ in SF$_6$ ambient. The employment of a high laser fluence led to the formation of a crater at the center of the irradiated area, with the appearance of various other surface topographies as we move towards the periphery of the crater, due to the Gaussian profile of the laser beam fluence. The major diameter of the elliptical crater is 870 μm and its depth 530 ± 20 μm. As the laser fluence decreases radially, the central crater is surrounded by a zone of silicon microspikes, similar to the ones generated with a lower laser fluence (Fig. 1), which is succeeded by a zone of microripples, followed by a zone of re-solidified silicon after melting [65].

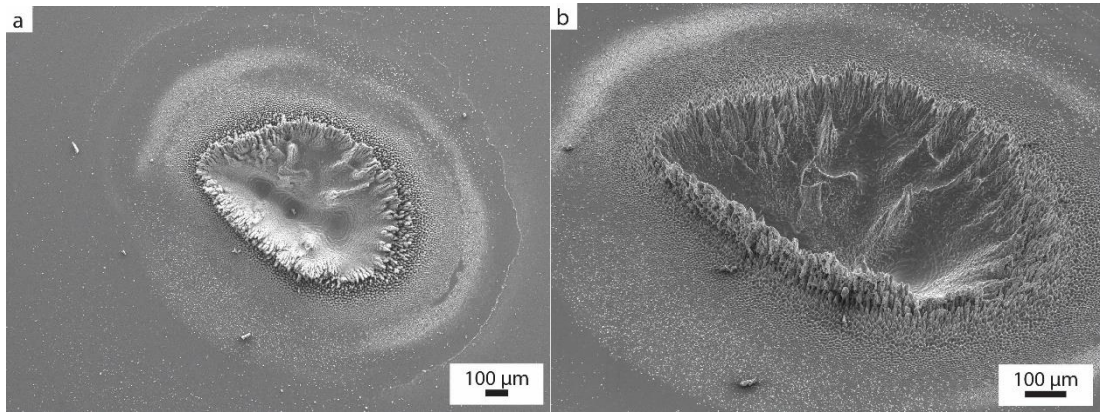

FIG. 3. SEM images at (a) top view and (b) side view (45°) of a crater formed by nanosecond-laser irradiation of silicon in $SF_6$ (fluence 2 J/cm$^2$).

Examination of cell attachment in and around these craters revealed the cells tend to concentrate more in the circumference of the crater (Fig. 4). Indeed, the cell population density appears slightly lower inside the crater rather than on the circumference and on the flat silicon area surrounding the crater.

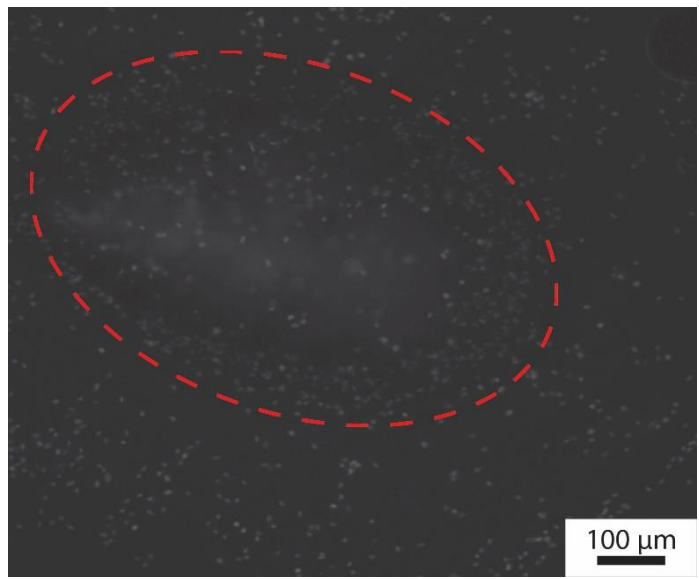

FIG. 4. Fluorescent micrograph of MDA-MB-231 cells in the area of the crater shown in Fig. 3. Dashed line indicates the crater border.

The crater effect on cell behavior becomes more pronounced for another crater morphology, which can be obtained by changing the environment in which laser irradiation takes place. Figure 5 shows an area on the silicon surface, which has been irradiated in vacuum, while the rest of the experimental conditions were kept identical to the conditions employed for the crater shown in Fig. 3 (1000 pulses, fluence 2 J/cm$^2$). For this crater, too, we observe the same effect as in Fig. 3, *i.e.*, an elliptical crater with major diameter 860 μm and depth 130 ± 20 μm at the center of the irradiated area, followed sequentially by microspikes, microripples, and finally molten and re-solidified silicon towards the edges. However, the fact that the crater shown in Fig. 5 was fabricated upon laser ablation in vacuum, results in a strikingly different morphology at the center, relatively to the crater fabricated in SF$_6$ ambient, as the former is shallower with vertical walls surrounding a central area which is decorated with microripples.

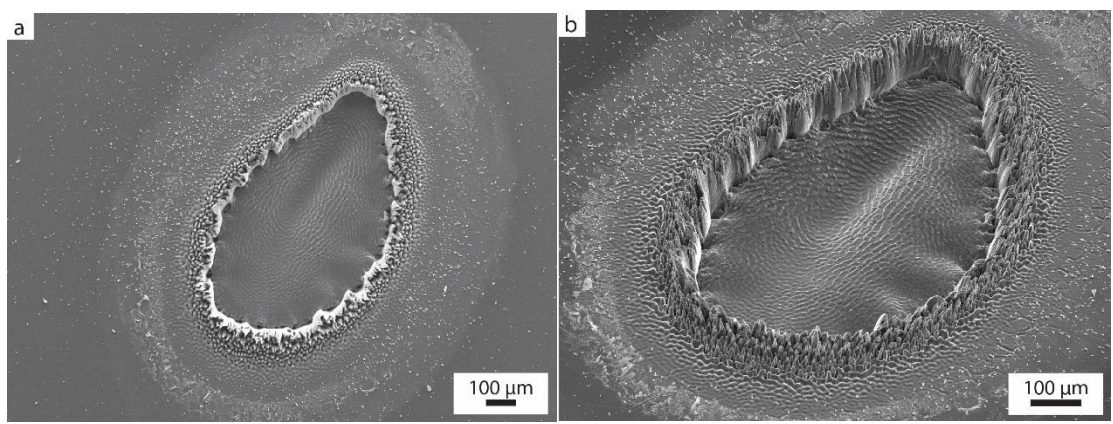

FIG. 5. SEM image at (a) top view and (b) side view (45º) of a crater formed by nanosecond-laser irradiation of silicon in vacuum (fluence 2 J/cm$^2$).

Figure 6 shows the behavior of MDA-MB-231 cells when seeded in the crater presented in Fig. 5. We observe that the cells prefer to adhere to specific sites around the crater, avoiding the zone with the microspikes and developing efficiently on a zone with sub-micron roughness, as we show below. The adhesion of cells was similar to the flat silicon surface and to the glass coverslip used as the positive control surface. This effect was further verified by quantitative analysis, which showed that 652 ± 15.63 (N=3) cells were located around the crater (on the microripple zone), whereas 600 ± 19.43 (N=3) cells were attached inside the crater. Taking into account the different areas of the crater and the microripple zone, the density of cells on the microripple zone is higher, indicating the preference of cells towards this topography.

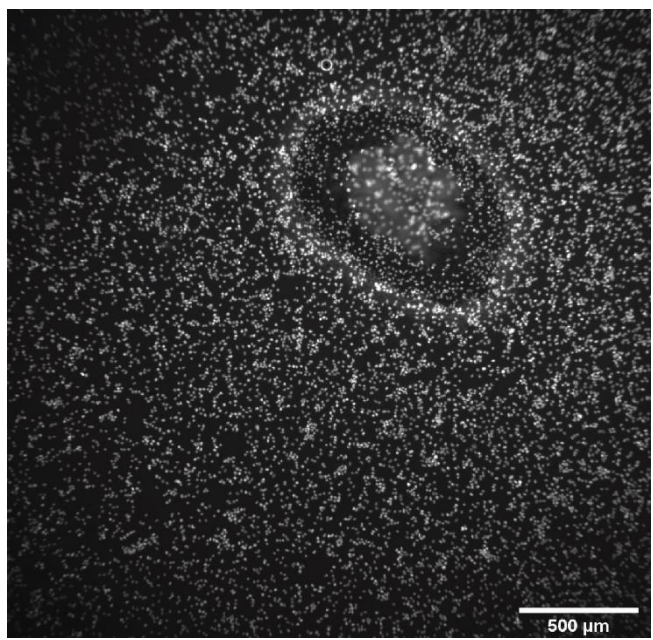

FIG. 6. Fluorescent micrograph of MDA-MB-231 cells on the area of the crater shown in Fig. 5.

We studied the behavior of the cells around the crater shown in Fig. 5 in more detail by creating a series of four such craters side-by-side, as shown in Fig. 7. The four craters present a similar morphology to the single crater shown in Fig. 5, with a central flat cavity, surrounded by a zone with microspikes, followed by a zone of microripples. The circumference of the craters is shown in more detail in Fig. 7c, which shows that the microspikes protrude above the surface of flat silicon, while the microripples have significantly smaller height than the microspikes and a smaller curvature. Fig. 7d shows a magnification of the microripples, where we observe the shape of the microripples and the fact that they are decorated with sub-micron roughness on the surface.

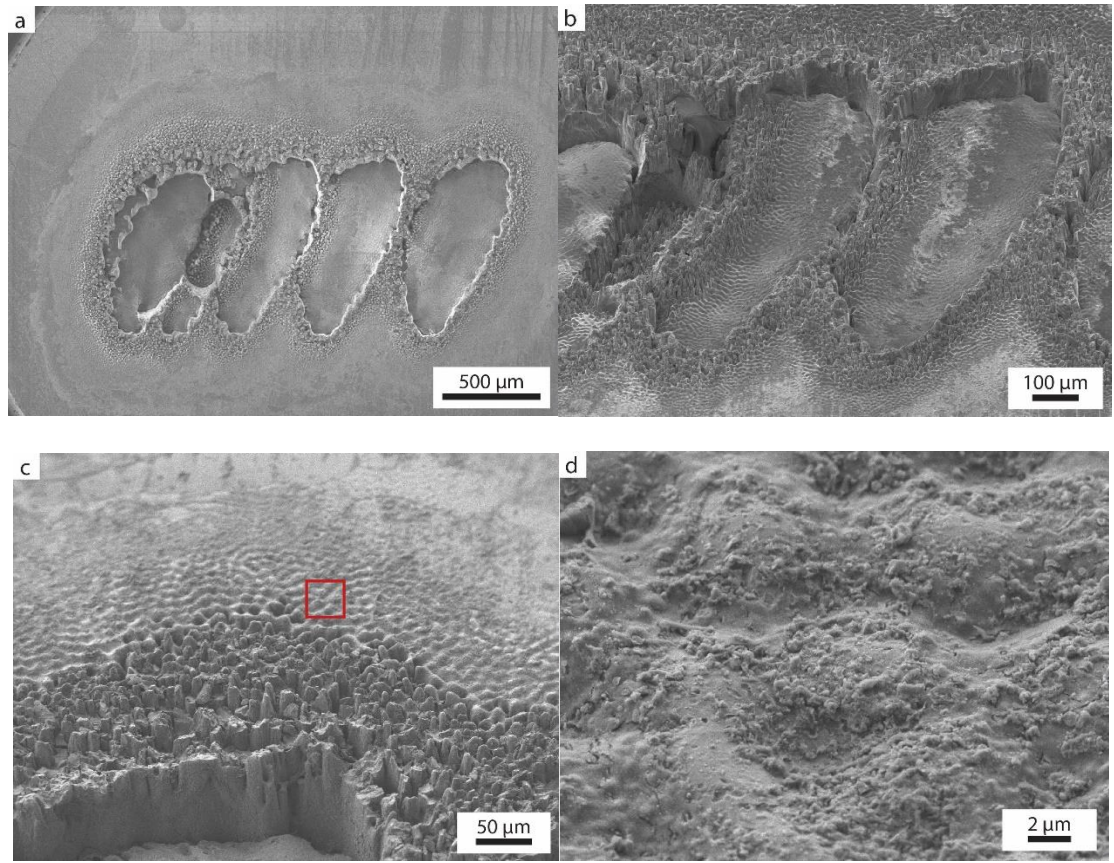

FIG. 7. SEM images at (a) top view and (b) side view (45º) of four adjacent craters formed by nanosecond-laser irradiation of silicon in vacuum (fluence 2 J/cm$^2$). (c) The circumference of the craters in more detail and (d) magnification of the area indicated by the red square in (c).

Cell adhesion on the series of four craters is demonstrated in Fig. 8, which shows again that the cells developed increased adherence capacity towards the areas with the microripples. Furthermore, the difference between microripples and conical microspikes in terms of cell preference becomes evident, as cells congregated in the rippled topography (indicated by the red rectangles), with a sub-micron surface roughness and a different aspect ratio than the microspikes, rather than the spiked one (indicated by the red circles). Similar to Fig. 6, quantification analysis showed that 994 ± 67.87 (N=3) cells adhered around the crater, while 691 ± 22.59 (N=3) cells localized inside the crater. The fact that different types of microstructures can affect cancer cell adherence, and consequently proliferation and growth, in various ways is further confirmed by the results obtained by other groups [66]. Therefore, the next step is to study cellular attachment on substrates with nanomorphology, as the above observations indicate that a substrate morphology exhibiting roughness at the nanoscale is significantly more favorable for cell adherence.

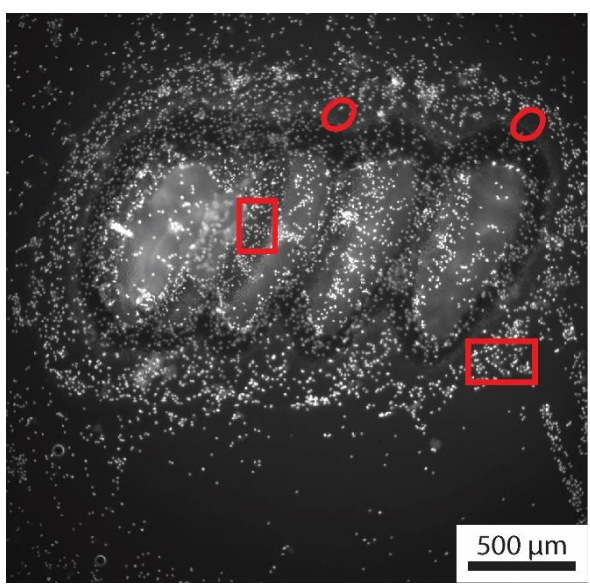

FIG. 8. Fluorescent micrograph of MDA-MB-231 cells in the area of the four craters shown in Fig. 7. Rectangles indicate the microripple area and circles the area with microspikes.

## B. MDA-MB-231 cell culturing on nanostructured substrates

We obtained nanostructured silicon substrates by employing a femtosecond laser system. Femtosecond laser irradiation of silicon in water, as described in the Experimental Section, resulted in the formation of columnar nanopillars on the surface, as shown in Fig. 9 [67, 68]. This is the result of ultrafast melting and interference effects, which occur only with femtosecond laser irradiation and not with nanosecond irradiation [69, 70]. The nanopillars were monolithically formed on silicon with a mean pillar diameter of 168 ± 33 nm, a mean height of 476 ± 68 nm, and a mean distance between neighboring nanopillars 257 ± 56 nm center-to-center.

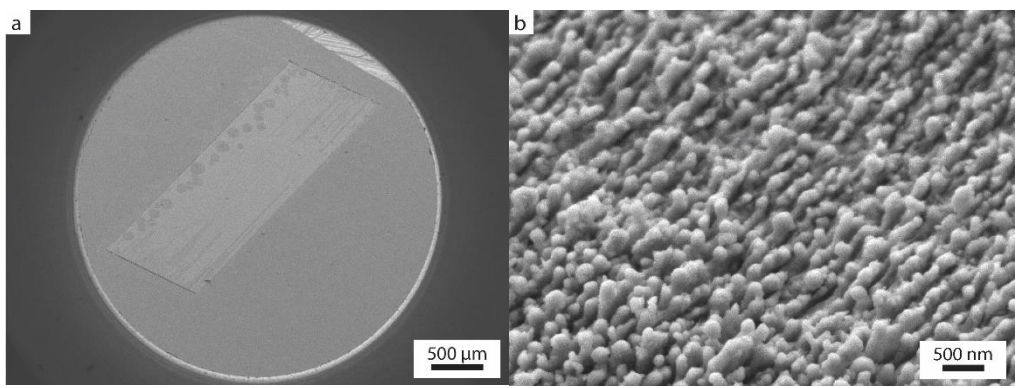

FIG. 9. SEM images of (a) a rectangular silicon area, processed with fs laser in water, and (b) higher magnification of the processed area at side view (45º).

Figure 10 shows the results of cell attachment on the nanostructured silicon surface. Two different rectangular areas of laser-processed silicon, similar to the one shown in Fig. 9a, co-exist on the same wafer and are surrounded by a flat silicon area. The MDA-MB-231 cells showed a remarkable preference for the nanostructured areas and accumulated on the nanopillars to a great extent. Specifically, 493 ± 3.844 (N=3) cells (Fig. 10a) and 465 ± 2.517 (N=3) cells (Fig. 10b) were attached to the nanostructured areas while 3 ± 1.155 (N=3) and 12 ± 2.082 (N=3) cells, respectively, were found on the flat areas aside. This is in stark contrast to the behavior of the cells on the microspikes shown in Fig. 1, which proved to be cell-repellent compared to the surrounding flat silicon area. Therefore, the nanoscale proves to be a naturally favorable topographical scale for MDA-MB-231 cell adherence. Nano-roughness is considered to be the closest to natural tissue morphology with a positive effect on cell adhesion, growth, and maturation. In human venous endothelial cells, it has been shown that increasing the roughness of the biomaterial surface at the nanometer scale can enhance cell adhesion and growth [22]. Mesenchymal MDA-MB-231 cells show maximal response to the influence of substrate morphology for cue dimensions close to the focal adhesion size [14].

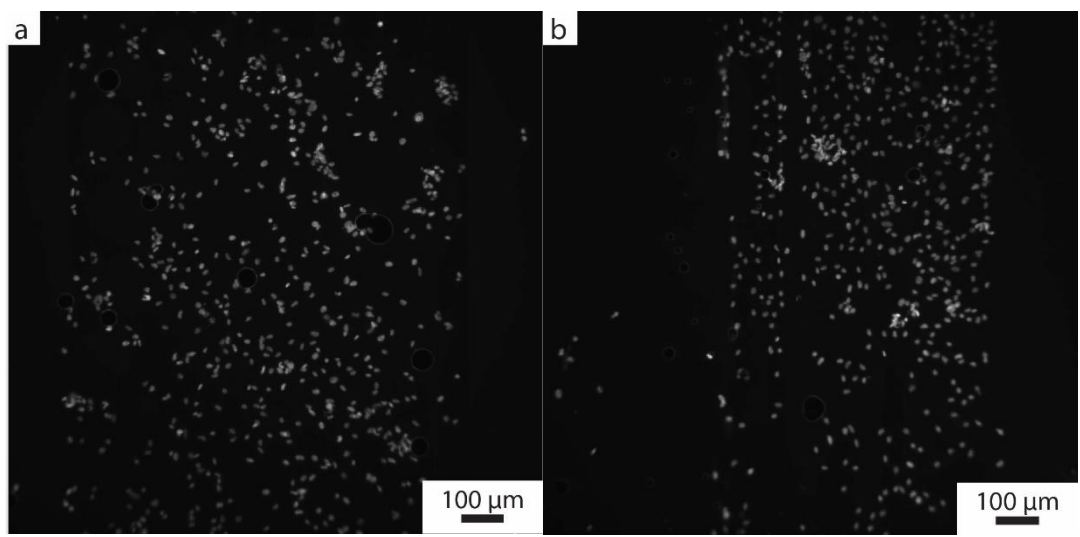

FIG. 10. Fluorescent micrographs of MDA-MB-231 cells on different rectangular areas of laser-nanostructured silicon, similar to the one shown in Fig. 9, surrounded by flat silicon.

## IV. SUMMARY AND CONCLUSIONS

Employing laser-patterning methods, we were able to modify the surface of silicon, inducing two different topographical scales, *i.e.*, microtopography and nanotopography, by varying the experimental parameters, most notably the laser pulse duration and the processing medium. Laser processing is a simple method to create micro/nanopatterned surfaces over large areas, without the need of a clean room. We used these substrates in order to compare the effect of topography on the adherence of MDA-MB-231 breast cancer cells in the absence of any other biochemical modification. The microtopographies we obtain by nanosecond laser irradiation, either conical microspikes or micro-craters, prove to be cell-repellent. On the other hand, the nanotopography we obtain by femtosecond laser irradiation, consisting of silicon nanopillars, is cell-attractive. The same observation holds for smooth microtopographies, such as microripples, decorated with sub-micron surface roughness, introduced with the use of nanosecond laser irradiation. The response of MDA-MB-231 cells appears maximum for substrate cue dimensions close to their focal adhesion size. Regulating cell attachment by inducing adhesion or repellence at specific surfaces will thus provide invaluable information about the proliferative and migrating behavior of breast cancer cells and will aid in designing novel treatment approaches. Gaining a deeper understanding of the topographical cues that govern TNBC adherence and elucidating cell responsiveness to mechanical signals, will enable further insights to be gained into prognostic markers, mechanically mediated metastasis pathways for therapeutic targets, and model systems required to advance cancer mechanobiology [71].

# ACKNOWLEDGMENTS

This research is co-financed by Greece and the European Union (European Social Fund – ESF) through the Operational Programme "Human Resources Development, Education and Lifelong Learning 2014 – 2020" in the context of the project "Novel studies of cell interactions on nanostructured surfaces" (MIS 5048213).

# AUTHOR DECLARATIONS

**Conflicts of Interest**

The authors have no conflicts to disclose.

**List of figure captions**

**FIG. 1.** Scanning electron micrographs (SEM) at (a) top view and (b) side view (45º) of laser-microstructured silicon with nanosecond laser beam in $SF_6$ (fluence 1 J/cm$^2$). (c) SEM image of flat silicon and (d) glass coverslip.

**FIG. 2.** Fluorescent micrographs of MDA-MB-231 cells on (a) glass coverslip, (b) flat silicon, (c) microstructured silicon as in Fig. 1, and (d) border between microstructured silicon (left side) and flat silicon (right side). Cell nuclei were stained with DAPI.

**FIG. 3.** SEM images at (a) top view and (b) side view (45º) of a crater formed by nanosecond-laser irradiation of silicon in $SF_6$ (fluence 2 J/cm$^2$).

**FIG. 4.** Fluorescent micrograph of MDA-MB-231 cells in the area of the crater shown in Fig. 3. Dashed line indicates the crater border.

**FIG. 5.** SEM image at (a) top view and (b) side view (45º) of a crater formed by nanosecond-laser irradiation of silicon in vacuum (fluence 2 J/cm$^2$).

**FIG. 6.** Fluorescent micrograph of MDA-MB-231 cells on the area of the crater shown in Fig. 5.

**FIG. 7.** SEM images at (a) top view and (b) side view (45º) of four adjacent craters formed by nanosecond-laser irradiation of silicon in vacuum (fluence 2 J/cm$^2$). (c) The circumference of the craters in more detail and (d) magnification of the area indicated by the red square in (c).

**FIG. 8.** Fluorescent micrograph of MDA-MB-231 cells in the area of the four craters shown in Fig. 7. Rectangles indicate the microripple area and circles the area with microspikes.

**FIG. 9.** SEM images of (a) a rectangular silicon area, processed with fs laser in water, and (b) higher magnification of the processed area at side view ($45^o$).

**FIG. 10.** Fluorescent micrographs of MDA-MB-231 cells on different rectangular areas of laser-nanostructured silicon, similar to the one shown in Fig. 9, surrounded by flat silicon.